\documentclass{elsart3p}
\usepackage{graphicx,amssymb}
\usepackage[square,numbers]{natbib}
\begin{document}
\begin{frontmatter}

\title{On the basic mechanism of Pixelized Photon Detectors}
\author[Tokyo]{H.~Otono\corauthref{cor}},~ 
\corauth[cor]{corresponding author.}
\ead{otono@icepp.s.u-tokyo.ac.jp}
\author[Tokyo]{H.~Oide},~ 
\author[ICEPP]{S.~Yamashita},~ 
\author[KEK]{T.~Yoshioka}

\address[Tokyo]{Department of Physics, School of Science, The University of Tokyo, 7-3-1 Hongo, Bunkyo-ku, Tokyo 113-0033, JAPAN}
\address[ICEPP]{International Center for Elementary Particle Physics, The University of Tokyo, 7-3-1 Hongo, Bunkyo-ku, Tokyo 113-0033, JAPAN}
\address[KEK]{Neutron Science Laboratory, High Energy Accelerator
Research Organization (KEK), 1-1 Oho, Tsukuba, Ibaraki 305-0801, JAPAN}

\begin{abstract}
  A Pixelized Photon Detector (PPD) is 
  a generic name for the semiconductor devices operated in the Geiger-mode,
  such as Silicon PhotoMultiplier and Multi-Pixel Photon Counter, 
  which has high photon counting capability.
  While the internal mechanisms of the PPD have been intensively studied 
  in recent years,
  the existing models do not include the avalanche process.
  We have simulated the multiplication and quenching of the 
  avalanche process and have succeeded in reproducing the output waveform of the PPD.
  Furthermore our model predicts the existence of dead-time in the PPD which
  has never been numerically predicted.
  For serching the dead-time, we also have developed waveform analysis method
  using deconvolution which has the potential to distinguish neibouring pulses
precisely.  
  In this paper, we discuss our improved model and waveform analysis method.
\end{abstract}

\end{frontmatter}

\section{Introduction}

A Pixelized Photon Detector (PPD) \cite{Amsler:2008zz} is a novel semiconductor photon sensor
which has a single photon level sensitivity and a linear relationship
between the gain and the operation voltage \cite{Renker:2006ay}.
Moreover the output waveform of the PPD has two component and
is not described by the typical single exponent function \cite{Piemonte:2007IEEE,Otono:2007PD07}.
While the internal mechanisms of the PPD have been intensively studied
in recent years,
the existing models incorporate artificial manipulation and
are nothing more than the phenomenology which explains above characteristics.
Thus, we have developed a new model based on semiconductor physics
and succeed in reproduction of the all characteristics.

 \begin{figure}[h]
  \begin{center}
   \includegraphics*[width=75mm]{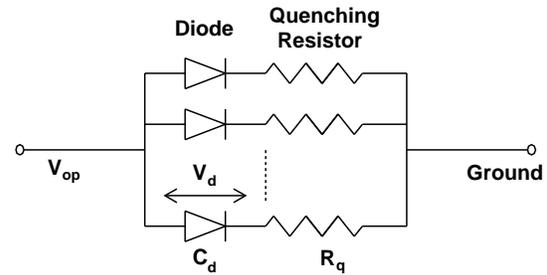}
  \end{center}
  \caption{A simple circuit of the PPD. Each pixel consists
  of a diode and a quenching resistor.}
  \label{fig:circuit}
 \end{figure}

A traditional model is shown in Figure \ref{fig:circuit}, where $V_{op}$ is the
operation voltage, $C_d$ the capacitance value of the diode and $R_q$ the quenching resistor value.
The main difference between our model and the previous ones is the
precondition of the applied voltage to the diode-part in the PPD
($V_{d}$).
The existing models fixed the minimum value of $V_d$ at the breakdown
voltage ($V_{br}$)
and the avalanche multiplication is forced to be terminated when $V_d$
reaches $V_{br}$.
On the contrary, we did not request such a constraint 
and the avalanche process is
terminated due to the internal physics: charge transportation, impact
ionization and circuit
equation \cite{Maes:1990Solid,Jacobi:2007Solid}.
On the other hand, $V_d$ may drops below $V_{br}$, and hence the gain linearity is not trivially determined.  

We have though calculated and we have in fact confirmed the gain linearity and 
the difference between $V_{min}$ and $V_{br}$, 
therefore the existence of significant dead time of the PPD
is predicted.

In this paper, we firstly introduce our model of the PPD and the
prediction of the dead time.
Next, we report the waveform analysis method which is effective to
search the dead time.

\section{A new model incorporating the avalanche process}
\begin{figure}[h]
  \begin{center}
    \includegraphics*[width=75mm]{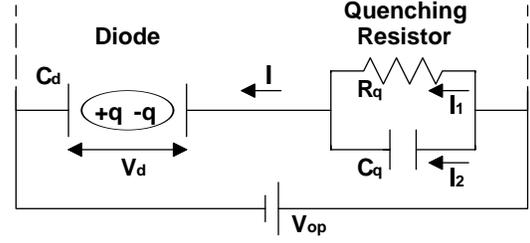}
  \end{center}
  \caption{A new equivalent circuit proposed in this paper.}
  \label{fig:circuit3}
\end{figure}

Figure \ref{fig:circuit3} shows our model and characterized by the
charge of created carriers
$q(t)$ in the diode which is time dependent in order to treat the
transient multiplication correctly \cite{Otono:2008NIMA}.
In this Figure, $I(t)$ means the observable current and $C_q$
describes a capacitance located between the diode-part and the
resistor-part, which has already reported by ITC-irst group \cite{Piemonte:2007IEEE,Piemonte:2007NIMA,Corsi:2007NIMA}.
Note that, $V_d(t)$ is determined by both
$q(t)$ and $I(t)$: $V_d(t)=V_{op}-\frac{q(t)-\int I(t)~dt}{C_d}$.
Using this formula, the circuit equation is calculated as follows:
\begin{equation}
\frac{d V_d(t)}{dt}=\frac{1}{R_q(C_d+C_q)}(V_{op}-V_d(t)-R_q\frac{dq(t)}{dt}).\\
\label{eqn:dv}
\end{equation}

The variation of the created carriers $\frac{dq(t)}{dt}$ is described by 
the densities of electrons and holes in the diode ($\rho_e$, $\rho_h$), 
the ionization rates ($\alpha_e$, $\alpha_h$), 
the drift velocities ($v_e$, $v_h$) and is expressed by:
\begin{equation}
  \frac{d q(t)}{dt}=e\sum_{i=e,h}\int{\rho_i}_{(x)}{\alpha_i}_{({V_d}_{(t)})}v_i~dx^3,
\label{eqn:dq}
\end{equation}
The ionization rate strongly depends on the electric field ($E(x)$) and 
is proportional to $E(x)\exp(-\frac{B}{E(x)})$, where 
$B$ is a constant.
We caluculated the $E(x)$ of the typical PPD \cite{Kagawa:2005IEICE} and obtained the almost uniform
$E(x)$ in the multiplication layer in the diode.
Thus, we assumed that $\alpha_e$ and $\alpha_h$ are the fuction of $V_d$.

Equations (\ref{eqn:dv}), (\ref{eqn:dq}) describe the evolution in the PPD
and can be sequencially solved.
Consequencely, the observable current is culculated from the following equation:

\begin{equation}
I(t)=C_d\frac{dV_d(t)}{dt}+\frac{dq(t)}{dt}.
\label{eqn:I}
\end{equation}

\begin{figure}[h]
  \begin{center}
    \includegraphics*[width=85mm]{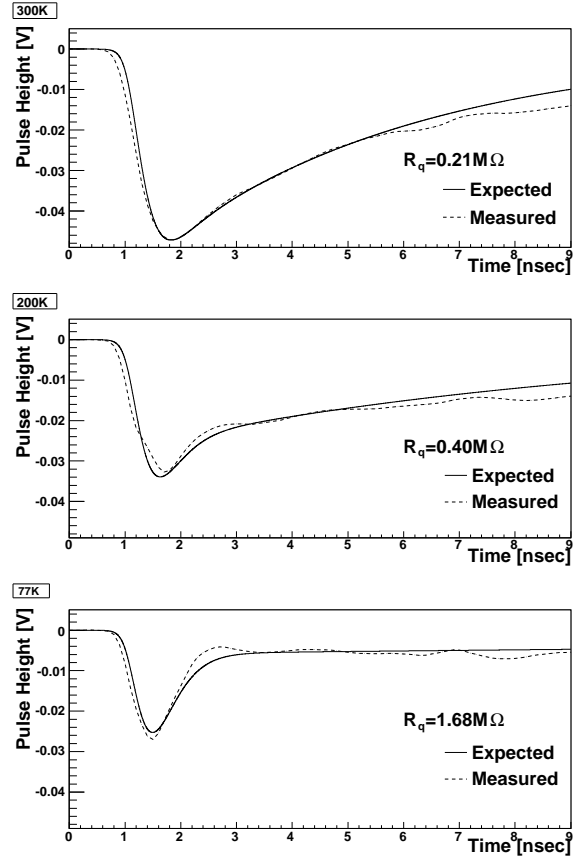}
  \end{center}
  \caption{The waveform expected from our proposed model and the measured 
waveform.}
  \label{fig:Final}
\end{figure}

Figure \ref{fig:Final} shows that the waveforms expected from equation (\ref{eqn:I}) (solid) agree well with the measured ones (dashed).
Note that, the transmission properties of the measurement system are already considered. 

\begin{figure}[h]
  \begin{center}
    \includegraphics*[width=85mm]{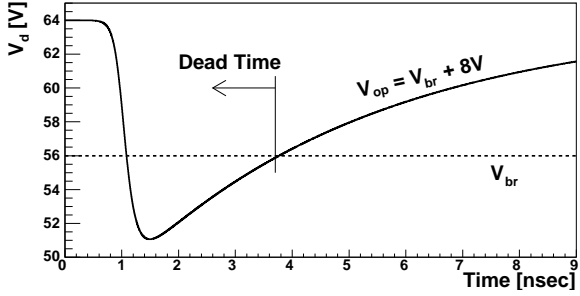}
  \end{center}
  \caption{The waveform expected from our proposed model and the measured 
waveform.}
  \label{fig:Deadtime}
\end{figure}

Figure \ref{fig:Deadtime} shows the time variation of $V_d$ and 
indicate that $V_d$ drops below $V_{br}$, therefore a few nano seconds
dead-time is expected.
We ignored the the carrier lifetimes and recombinations which decrease 
the number of the carriers.
If the effects are taken into account, the dead-time seems to be less than
predicted, however it will still be significant. 

\section{Waveform analysis}

For the precise measurement of average waveform of PPD, or for the study of crosstalk, afterpulsing, recovery, etc., it is effective to use and analyze waveforms. We are using a digital oscilloscope and measuring the waveform for studying basic characteristics of PPDs. Consequently, waveform analysis algorithm is important. Here we introduce the deconvolution method we are developing.

If the response function $h(t)$ to a particular impulse is known, {\it deconvolution} using Fourier analysis is effective to detect pulses precisely for linear response systems. Here we describe the outline of this algorithm.

Suppose that the observed waveform is written as follows:
\begin{eqnarray}
\label{eq1}
y(t) &=& \int dt' \sum_{j}a_{j}\delta(t'-t_{j})h(t-t') + n(t)\nonumber\\
	 &=& \int dt' d(t')h(t-t') + n(t)~,
\end{eqnarray}
where $d(t)\equiv\sum_{j}a_{j}\delta(t-t_{j})$ denotes the array of impulses, which is referred as {\it comb function}, and $n(t)$ denotes the noise component of the observed waveform. Following this assumption, out problem is to estimate the comb function as precise as possible and to find and measure the position and the size of peaks $\{a_{j},t_{j}\}$ from the estimated comb function. In the frequency domain, Eq.\ref{eq1} is represented as
\begin{equation}
\label{eq2}
Y(\omega) = D(\omega)H(\omega) + N(\omega)~,
\end{equation}
where $Y(\omega), D(\omega)$, and $H(\omega)$ are Fourier transformations of $y(t), d(t)$, and $h(t)$, respectively. Considering the ideal case that the noise component is neglected, Eq.\ref{eq2} is replaced to
\begin{equation}
Y(\omega)=D(\omega)H(\omega)~.
\end{equation}
Dividing $Y(\omega)$ by $H(\omega)$, $D(\omega)$ is obtained, and the problem is solved as this is equivalent to the comb function $d(t)$:
\begin{eqnarray}
d(t) &=& {\cal F}^{-1}\left[ \frac{1}{H(\omega)}Y(\omega) \right] = {\cal F}^{-1}\left[H_{D}(\omega)Y(\omega)\right]\nonumber\\
 &=& (h_{D}\otimes y)(t)
\end{eqnarray}
where $H_{D}(\omega)\equiv H(\omega)^{-1}$ and the symbol $\otimes$ denotes the convolution.

Next we cope with the real calse. In the real case that the noise component exists, the deconvolution method cannot be used directly because the power spectrum of the response function $|H(\omega)|^{2}$ is typically large in lower frequency region and is small in higher frequency region. Consequently the deconvolution filter enhances the high frequency region of the observed waveform $y(t)$. As the power spectrum of the noise component is roughly flat in frequency, this manner of the deconvolution filter emphasizes the noise component, and as a result the deconvoluted waveform oscillates keenly. This is not the desired one.

\begin{figure}[h]
  \begin{center}
    \includegraphics*[width=85mm]{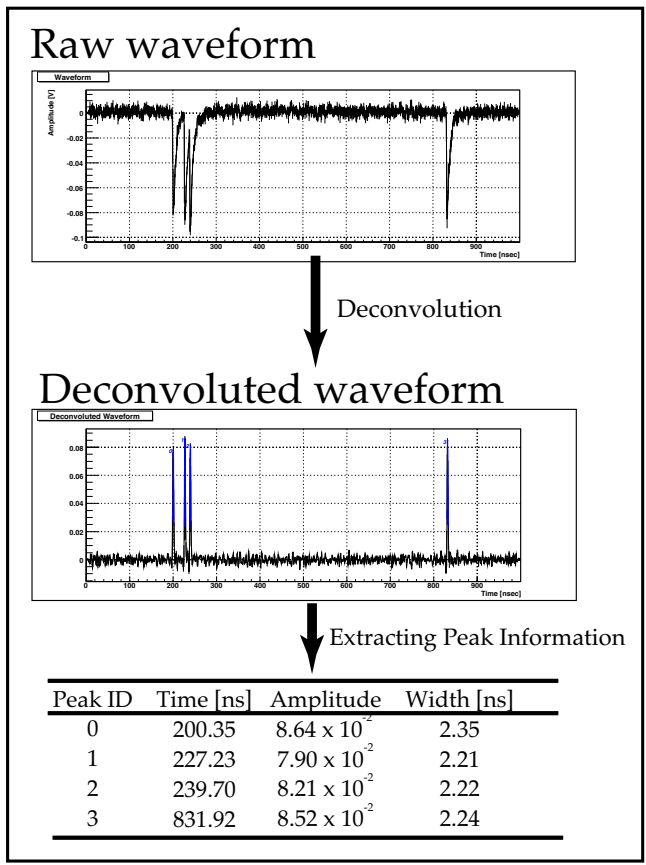}
  \end{center}
  \caption{The scheme of the waveform analysis using deconvolution method.}
  \label{fig:deconv}
\end{figure}

To avoid this problem, couple of the deconvolution filter with some low-pass filters (LPFs) is effective. Several types of LPFs are selectable e.g. Wiener optimal filter, at present we adopt the proper LPF:
\begin{equation}
LPF(\omega) = \frac{1}{\sqrt{1+(\omega/\omega_{c})^{2}}}
\end{equation}
where $\omega_{c}$ is the cutoff frequency that is determined from the relation of the time resolution and the confidence level of the peak detection. The higher the cutoff frequency the lower the SNR, and the lower the lower the time resolution. Using this deconvolution method we can split pile-up neighboring pulses up to 2 [nsec] intervals. Figure \ref{fig:deconv} shows the scheme of the deconvolution method.

\section{Conclusion and future prospects}
We have developed a new model of Pixelized Photon detectors and have succeeded 
in reproducting all characteristics, such as spike component in the waveforms 
and linearity relationship between the gain and the bias voltage.
Furthermore we our model predicts the existence of dead-time in the PPD 
which has not been numerically predicted.

Our waveform analysis method has the potential to distinguish neighbouring pulses upto about 2[nsec] interval and to determine the amplitude of each pulse precisely. Applying this method, various precise analyses are possible, e.g. measuring the recovery process to search dead-time, or inspecting the frequency distribution of afterpulsing, etc.

Now, the measurement of the dead-time and the verification of our model of PPD is on going.
We are also challenging to study the characteristics of PPDs using TCAD semiconductor simulation technology. This study will lead to strategic developments of PPDs, especially in developing optimized PPDs for applied experiments in the future.

\section{Acknowledgments}
The authors wish to express our deep appreciation to the Hamamatsu Photonics K.K. and the KEK-DTP
photon-sensor group members for their helpful discussions and suggestions.
This work was supported by Grant-in-Aid for JSPS Fellows $20\cdot4439$
and by Grant-in-Aid for Exploratory Research $20654021$
from the Japan Society for the Promotion of Science (JSPS).


\end{document}